\documentclass[prd,superscriptaddress,nofootinbib,a4paper,preprintnumbers]{revtex4} 

\usepackage{graphicx}
\usepackage{dcolumn}
\usepackage{bm}
\usepackage{latexsym}
\usepackage{amsfonts}
\usepackage{amssymb}
\usepackage{amsmath}
\usepackage{bbold}
\usepackage{ulem}
\usepackage[dvipsnames]{xcolor}
\usepackage[utf8]{inputenc} 
\usepackage{xspace}
\usepackage{ulem}
\newcommand{\Mpl}{{M_{\rm pl}\xspace}}
\allowdisplaybreaks[1]

\newcommand{\hh}{\mathfrak{h}}

\begin{document}

\date{\today}
\title{Gravitational wave interactions in $\Lambda_3$ models of dark energy}

\author{A. Emir G\"umr\"uk\c{c}\"uo\u{g}lu}
\affiliation{Institute of Cosmology and Gravitation, University of Portsmouth\\ Dennis Sciama
	Building, Portsmouth PO1 3FX, United Kingdom}

\author{Kazuya Koyama}
\affiliation{Institute of Cosmology and Gravitation, University of Portsmouth\\ Dennis Sciama
	Building, Portsmouth PO1 3FX, United Kingdom}

\date{\today}

\begin{abstract}
We argue that cubic order interactions between two scalar gravitons and one tensor graviton are ubiquitous in models of dark energy where the strong coupling scale is $\Lambda_3$. These interactions can potentially provide efficient decay channels for gravitational waves. They can also lead to gradient instabilities of the scalar perturbations in the presence of large amplitude gravitational waves, e.g. those detected by LIGO/Virgo. In contrast with models in scalar-tensor theories, there is an infinite number of higher order interactions in generic $\Lambda_3$ models, which make it difficult to predict the fate of these instabilities inferred from cubic order interactions.

\end{abstract}

\maketitle

\section{Introduction}
The concordance model of cosmology introduces a cosmological constant as the sole origin of late-time accelerated expansion. This approach is successful in providing the best fit to cosmological data, but unfortunately, cosmological constant is a particularly elusive model that does not provide many observable handles to allow probing it directly. Although alternative models to cosmological constant typically aim to resolve the fine-tuning problem, they also allow us to quantify how close the universe is to the concordance cosmology. Such alternatives usually contain new degrees of freedom that generate observable signals, e.g. a time dependent dark energy equation of state or a characteristic structure formation profile (see e.g. \cite{Koyama:2015vza} for a review).

In the last decade, dark energy models in modified gravity theories attracted significant interest. These theories typically have one or more degrees of freedom that participate in gravitational interactions, while providing an accelerated expansion without a vacuum energy. The simplest category of such theoretical constructions is obtained by introducing a single scalar field $\phi$ and described in the framework of the scalar-tensor theory class. The dark energy models in this class have some common features: they are effective field theories with a relatively low strong coupling scale compared to GR at $\Lambda_3 = (H_0^2\Mpl)^{1/3} \sim 10^{-22}{\rm GeV}$, where $H_0 \sim 10^{-33} {\rm eV}$ is the value of the Hubble rate today. Moreover, the scalar graviton often has a direct coupling to the matter sector which requires a mechanism to screen it from local tests of gravity and other observations that have confirmed the validity of the strong equivalence principle. 

Despite the rich observational possibilities of scalar-tensor dark energy models, direct detection of gravitational waves can impose severe constraints on the model parameters. However, the interpretation of these constraints may be obscured at energies around the strong coupling scale $\Lambda_3$, which is coincidentally close to the momentum scale of LIGO gravitational waves \cite{deRham:2018red}. Assuming that the effective field theory is still valid, there are several implications of LIGO\footnote{https://www.ligo.caltech.edu/}/Virgo\footnote{https://www.virgo-gw.eu/} observations. Foremost, the gravitational wave signal GW170817 from the binary neutron star merger with electromagnetic counterpart GRB 170817A indicates that tensor waves propagate at the speed of light to an accuracy of one part in $10^{15}$ \cite{LIGOScientific:2017zic}, potentially ruling out models that predict tensors with non-luminal speeds \cite{Ezquiaga:2017ekz, Creminelli:2017sry, Sakstein:2017xjx, Baker:2017hug}. Another implication, which is the focus of the present paper, involves the effects arising from interactions between the tensor perturbation $h_{ij}$ and the dark energy scalar perturbation $\Pi$ \cite{Creminelli:2018xsv, Creminelli:2019kjy, Creminelli:2019nok}. As a demonstration, let us consider 3-point tensor-scalar-scalar interactions. In the presence of a large tensor wave with frequency $\omega$, the scalar graviton perturbation $\Pi$ is shifted away from its vacuum configuration and is assigned the same spatial and temporal gradient as the gravitational wave source, i.e. $\partial\Pi/\Pi \sim \partial h/h\sim \omega$. In this arrangement, the tensor-scalar-scalar interactions now can pose some threats on the stability of the scalar perturbations. These interactions are formally
\begin{equation}
\mathcal{L}^{(3)} \ni \frac{a_0}{\Mpl} h_{ij}\partial^i\Pi\,\partial^j\Pi+ \frac{a_1}{\Lambda_2^2} \dot{h}_{ij}\partial^i\Pi\,\partial^j\Pi + \frac{a_2}{\Lambda_3^3} \ddot{h}_{ij}\partial^i\Pi\,\partial^j\Pi + \frac{a_3}{\Lambda_3^3}\partial_k\partial^kh_{ij}\partial^i\Pi\,\partial^j\Pi \,,
\label{eq:formal-interactions}
\end{equation}
where the perturbations are canonically normalised, $\partial_i$ denotes a partial derivative with respect to spatial coordinates and $\Lambda_n\equiv \left(\Mpl m^{n-1}\right)^{1/n}$. Here, the meaning of $m$ depends on the context, but for dark energy models we typically have $m\sim H_0$. 

In Eq.\eqref{eq:formal-interactions}, the $a_0 \,h_{ij}\,\partial^i\Pi\partial^j\Pi$ term is the first correction to the quadratic scalar gradient term $\partial_i\Pi\partial^i\Pi$ and it can get contributions from various interaction terms in the scalar-tensor class. However, since it lacks the enhancement due to the gravitational wave frequency, it is typically suppressed compared to the other interactions.

In scalar-tensor theories, the $a_1\dot{h}_{ij}\,\partial^i\Pi\,\partial^j\Pi$ interaction arises from terms in the Lagrangian that contain second covariant derivatives of the scalar $\phi$ of the form $f(\nabla\phi)\nabla\nabla\phi$, and has the same origin as the kinetic braiding effect \cite{Deffayet:2010qz} at linear order. In the presence of LIGO-scale gravitational waves, this term can dominate over the usual (quadratic) gradient term for the scalar. Being a cubic order term, it causes the complete gradient to have a spacetime dependent sign, i.e. removes the lower bound from gradient energy, hence is the source of a classical instability \cite{Creminelli:2019kjy}.

Finally, the $a_2\ddot{h}_{ij}\partial^i\Pi\,\partial^j\Pi$ term arises from higher covariant derivatives of $\phi$, common in Beyond-Horndeski \cite{Gleyzes:2014qga} and DHOST theories \cite{BenAchour:2016fzp}. The interaction is enhanced by two powers of the gravitational wave frequency and is responsible for the decay of tensor modes into the dark energy scalar \cite{Creminelli:2018xsv, Creminelli:2019nok}. 
Note that this decay is a \v{C}erenkov type process: if the sound speed of the dark-energy scalar is super-luminal, the decay is kinematically forbidden. The $a_2$ term also causes the effective gradient of the scalar mode to be unbounded, triggering a classical instability even more efficient than the one generated by the $a_1$ term \cite{Creminelli:2019kjy}. The final term in Eq.\eqref{eq:formal-interactions}, i.e. $a_3\partial_k\partial^k h_{ij}\, \partial^i\Pi\partial^j\Pi$, acts at the same scale as the $a_2$ term (although not necessarily at the same time).
It can originate from quartic and quintic galileon terms around a Minkowski background \cite{Koyama:2013paa}. It is also reasonable to expect it to appear in the most general DHOST theories.
If the coefficients $a_2$ and $a_3$ are independent, one might be tempted to tune these against each other to cancel the effect of the interactions in order to obtain a stable cosmology with gravitational waves. However, in a cosmological setup, these coefficients are generically time dependent, so unless there is a theory-specific reason, such a cancellation is difficult to justify. We therefore assume that $a_3$ term would also contribute to the instability.

The interactions we discussed up to now arise in scalar-tensor theories studied in 
\cite{Creminelli:2018xsv,Creminelli:2019kjy, Creminelli:2019nok}. However, in a more general $\Lambda_3$ effective theory, one can have higher-order interactions formally of the type
\begin{equation}
\mathcal{L}^{(n)} \ni \frac{1}{\Lambda_3^{3(n-2)}}\,h \,(\partial\partial \Pi )^{n-1}\,,
\label{eq:high-order-formal}
\end{equation}
where, $\partial$ without indices here denotes either a spatial or time derivative.
Due to the additional derivatives, these terms are enhanced further by higher powers of the frequency. Depending on the strength of the term that sources the scalar graviton, we may lose control of the perturbative expansion and the system may become strongly coupled. It is also possible that these non-linear interactions prevent the growth in $\Pi$, avoiding instabilities as well as the decay of the tensor modes \`a la Vainshtein. In some theories with Vainshtein mechanism, the screening is partially due to the interactions between the Newtonian potential and the scalar field: in a regime where there is a sizeable gravitational potential, the coupling between the potential and the scalar becomes strong, eventually leading to an effective decoupling of the scalar field. Thus from covariance, it would not be surprising to encounter similar interactions between the tensor mode and the scalar field \eqref{eq:high-order-formal}. These interactions are weak in a general cosmology context, but in the presence of a tensor source can have dramatic effects. Although screening of scalar-tensor interactions are possibility, confirming this scenario requires a non-perturbative approach, which is beyond the scope of this paper. However we can at least determine whether a strong coupling is triggered in the presence of LIGO gravitational waves.

The series of $h\,(\partial\partial \Pi )^{n-1}$ interactions are actually present in scalar-tensor theories but are truncated at a finite order. For instance, general Horndeski class has the interaction of the form \eqref{eq:high-order-formal} up to $n=4$ around the Minkowski spacetime \cite{Koyama:2013paa}, while for the cases studied in Refs.\cite{Creminelli:2018xsv,Creminelli:2019kjy, Creminelli:2019nok}, the truncation is at cubic order $n=3$. However, in a generic $\Lambda_3$ theory, the situation is different. To be specific, in massive gravity, due to the square-root nature of the graviton potential, one has an infinite series of interactions. This opens up the possibility of a strong coupling.  

In this article, we attempt to quantify the above arguments in the framework of a $\Lambda_3$ effective field theory that allows a larger set of interactions than the scalar-tensor class. To this aim, we consider generalised massive gravity (GMG) \cite{deRham:2014lqa,deRham:2014gla}, which is a special example in the general massive spin--2 theory class \cite{Gumrukcuoglu:2020utx}. This is an extension of de Rham-Gabadadze-Tolley (dRGT) massive gravity \cite{deRham:2010kj} with stable cosmologies \cite{Kenna-Allison:2019tbu} and Vainshtein mechanism \cite{Gumrukcuoglu:2021gua}, while yielding observational signatures similar to, but distinguishable from, scalar-tensor models \cite{Kenna-Allison:2020egn}.

The paper is organised as follows. In Sec.II, we review the cosmology of generalised massive gravity, then in Sec.III we present interaction terms corresponding to tensor-scalar-scalar, tensor-tensor-scalar and higher order one-tensor vertices. In Sec.IV, we investigate the implications of these interactions in the presence of a gravitational wave of LIGO scale. We conclude with Section V, where we discuss and summarise our results.

\section{Cosmology in generalised massive gravity}

In this Section, we review the cosmological solutions in generalised massive gravity theory. We also introduce perturbations and calculate the free Lagrangian for cosmological perturbations. In particular, we derive the three constraint equations for perturbations which will become essential in obtaining the interactions among the dynamical degrees of freedom in the next Section.

\subsection{The theory and the background cosmology}

The action we use consists of the Einstein-Hilbert term, generalised dRGT terms and a k-essence field $\zeta$ that represents the matter sector \cite{deRham:2014gla}:
\begin{equation}
S = \int d^4x\,\sqrt{-g}\,\left[\frac{\Mpl^2}{2}\,R +\Mpl^2m^2\,\sum_{i=0}^{4}\,\alpha_i(\Phi^a\Phi_a)\, e_i\left(\mathbb{1}-\sqrt{g^{-1}f}\right) + P(X)
\right]\,,
\end{equation}
where $\Phi^a$ are four scalar fields which are responsible for the breaking of the general coordinate invariance through the mass terms constructed out of $e_i$, the elementary symmetric polynomials.
For the time being, we also include a k-essence type matter field as a place-holder, with canonical kinetic term defined as $X \equiv -\frac12\,\partial_\mu\zeta\partial^\mu\zeta$.

For the background configuration, we use a variant of the non-Lorentz invariant gauge \cite{Gumrukcuoglu:2011zh, deRham:2014lqa} and adopt the zero curvature limit \cite{deRham:2014gla}. The resulting fiducial metric $f_{\mu\nu} \equiv \eta_{ab}\partial_\mu \phi^a\partial_\nu\phi^b$ is given by \cite{Gumrukcuoglu:2021gua},
\begin{align}
f_{00} &= 
-\left[  \chi'(t+\Pi^0) (1+\dot\Pi^0)\right]^2+\alpha^2\,\partial^i\dot\Pi^L\,\partial_i \dot\Pi^L
\,,\nonumber\\
f_{0i} &= 
-\left[  \chi'(t+\Pi^0)\right]^2 (1+\dot\Pi^0) \,\partial_i\Pi^0+\alpha^2\,\left[\partial_i\dot\Pi^L+\partial^k\dot\Pi^L\,\partial_k\partial_i\Pi^L\right]
\,,\nonumber\\
f_{ij} &= -\left[  \chi'(t+\Pi^0)\right]^2 \partial_i\Pi^0\,\partial_j\Pi^0 + \alpha^2\left(\delta_{ij}+2\,\partial_i\partial_j\Pi^L +\partial_i\partial^k\Pi^L\,\partial_k\partial_j\Pi^L\right)
\,,
\label{eq:fiducial_metric}
\end{align}
where $\Pi^0$ corresponds to temporal St\"uckelberg field perturbation, while $\Pi^L$ is the spatial longitudinal St\"uckelberg field perturbation. For convenience, we also introduced the constant $\alpha$ which keeps track of the normalisation of the spatial coordinates. 

In GMG, the mass parameters are promoted to functions that depend on the norm of the St\"uckelberg fields, which in the zero curvature limit evaluates to \cite{Gumrukcuoglu:2021gua}
\begin{equation}
 \Phi^a\Phi_a = \chi(t+\Pi^0)\,.
 \label{eq:PhiaPhia}
\end{equation}
For the physical metric, we introduce scalar and tensor perturbations in the spatially flat gauge:
\begin{equation}
 ds^2 = -N^2(1+2\,\Phi)dt^2 + 2\,N\,a\,\partial_i B dt \,dx^i + 
 a^2\left(\delta_{ij} + h_{ij}
 \right)dx^idx^j\,.
 \label{eq:metric-decomposed}
\end{equation}

Before discussing the perturbations, let us first look at the background equations of motion. First order variation of the action in the flat scaling limit, then fixing $N=1$, gives the background equations of motion as \cite{Gumrukcuoglu:2021gua}:
\begin{align}
 3\, H^2 =& m^2\,L + \frac{\rho}{\Mpl^2}\,,\nonumber\\
 2\,\dot{H} =& m^2(r-1)\,J\,\xi - \frac{\rho+P}{\Mpl^2}\,,\nonumber\\
 3\,H\,J =& L_\chi\,,\nonumber\\
 \dot{\rho} =& -3\,H\,(\rho+P)\,,
 \label{eq:background-eom}
\end{align}
where $\xi \equiv\frac{\alpha}{a}$, $r \equiv \frac{\dot{\chi}}{\xi}$, $H \equiv \frac{\dot{a}}a$, while $P$ and $\rho$ are the analogues of pressure and energy density for the k-essence matter. In the above,
we defined the following to represent the mass functions:\footnote{Note that the definition of these functions is slightly modified with respect to Ref.\cite{Kenna-Allison:2020egn}, where $L$ and $J$ are the same, but $Q$ and $G$ are different. For instance, the dRGT tensor mass in Ref.\cite{Kenna-Allison:2020egn} corresponds to $\Gamma \to \xi\,[J + Q\,\xi\,(r-1)]$.}
\begin{align}
L\equiv & -\alpha_0(\chi)+(3\,\xi-4)\alpha_1(\chi)-3\,(\xi-2)(\xi-1)\alpha_2(\chi)+(\xi-4)(\xi-1)^2\alpha_3(\chi)+(\xi-1)^4\alpha_4(\chi)\,,
\nonumber\\
J\equiv  \frac{1}{3}\,\frac{\partial L}{\partial \xi} =&\alpha_1(\chi) + (3-2\,\xi)\alpha_2(\chi)+(\xi-3)(\xi-1)\alpha_3(\chi)+(\xi-1)^2\alpha_4(\chi)\,,
\nonumber\\
Q\equiv  \frac{1}{2}\,\frac{\partial J}{\partial \xi}=&
-\alpha_2(\chi)+(\xi-2)\alpha_3(\chi)+(\xi-1)\alpha_4(\chi)\,,
\nonumber\\
G\equiv  \frac{\partial Q}{\partial \xi}=& \alpha_3(\chi)+\alpha_4(\chi)\,,
\label{eq:defLJQG}
\end{align}
with the derivatives defined as $J_\chi \equiv \frac{\partial J}{\partial\chi}$, $J_{\chi\chi} \equiv \frac{\partial^2 J}{\partial\chi^2}$ and similarly for the other functions. 

The system of equations \eqref{eq:background-eom} is closed and the last line can be derived via contracted Bianchi identities, or equivalently, from energy conservation.

\subsection{Quadratic action, constraints and canonical normalisation}
We now move on to the action quadratic in perturbations,
\begin{align}
 \delta^2S = \frac{\Mpl^2}{2}\,\int dt\,a^3 d^3x\,\mathcal{L}^{(2)}\,,
\end{align}
where, using the decompositions \eqref{eq:fiducial_metric} and \eqref{eq:metric-decomposed}, we have
\begin{align}
\mathcal{L}^{(2)} =&
\frac{m^2J\,\xi}{1+r}\left[
\partial_iB\partial^iB+\frac{r^2}{a^2}\,\partial_i\Pi^0\partial^i\Pi^0
+a^2\,\partial_i\dot\Pi^L\partial^i\dot\Pi^L
-2\,r\,\Pi^0\partial^2\dot\Pi^L
-\frac{2\,r^2}{a}\,\Pi^0\,\partial^2B
+2\,a\,\dot\Pi^L\,\partial^2B
\right]\nonumber\\
&+6\,m^2\,J\,\xi\,r\,\Pi^0\,\left[-H\,\Phi + \frac{1}{4}\,\left(\frac{\rho+P}{\Mpl^2}-m^2J\,\xi(r-1)\right)\Pi^0\right]
-m^2r\,\xi^2(J_\chi-2\,H\,Q)\,\left(2\,\partial^2\Pi^L+3\,H\,\Pi^0\right)\,\Pi^0
\nonumber\\
&-2\,m^2J\,\xi\,\Phi\,\partial^2\Pi^L
-\frac{4\,H}{a}\,\Phi\,\partial^2B
+\left(\frac{\rho+P}{\Mpl^2\,c_\zeta^2}-6\,H^2\right)\,\Phi^2\nonumber\\
&+\frac{(\rho+P)}{\Mpl^2\,\dot\zeta^2}\left[
\frac{1}{c_\zeta^2}\,\left(\delta\dot\zeta-2\,\dot\zeta\,\Phi\right)\delta\dot\zeta-\frac{1}{a^2}\,\partial_i\delta\zeta\,\partial^i\delta\zeta
+\frac{2}{a}\,\dot\zeta\,\delta\zeta\,\partial^2B
\right]
\,,
\label{eq:Lagrangian}
\end{align}
where $\delta\zeta$ denotes the perturbation of the matter field, while $c_\zeta$ is the sound-speed for the equivalent fluid. We also defined the spatial Laplacian $\partial^2 \equiv \partial^i\partial_i$.
In the above form, we observe that $\Phi$, $B$ and $\Pi^0$ appear without any time derivatives, thus are non-dynamical. The equation of motion for $B$ can be integrated twice to give
\begin{equation}
-\frac{m^2\,a\,J\,\xi}{2\,(r+1)}\,\left(B - a\,\dot\Pi^L+\frac{r^2}{a}\,\Pi^0\right)+\frac{\rho+P}{2\,\Mpl^2\,\dot\zeta}\,\delta\zeta - H\,\Phi=0\,,
\label{eq:Bcons}
\end{equation}
which can be solved algebraically for $B$ in terms of other perturbations. For the remaining two non-dynamical modes, the situation is more complicated. The equation of motion for $\Phi$ is obtained as:
\begin{equation}
\left(\frac{\rho+P}{c_\zeta^2\,\Mpl^2}-6\,H^2\right)\Phi - \frac{\rho+P}{c_\zeta^2\,\dot\zeta\,\Mpl^2}\,\delta\dot\zeta - 3\,m^2H\,r\,J\,\xi\,\Pi^0 - \partial^2\left(
\frac{2\,H}{a}\,B +m^2J\,\xi\,\Pi^L
\right)=0\,,
\label{eq:Fcons}
\end{equation}
while the equation for $\Pi^0$ is:
\begin{align}
-\frac{1}{2}\,\left(m^2\,J\,\xi\,(r-1)-\frac{\rho+P}{\Mpl^2}\right)\,\Pi^0-H\,\Phi - \frac{H\,\xi\,(J_\chi-2\,H\,Q)}{J}\,\Pi^0~~~~~~~~~~~~&
\nonumber\\
-\frac{1}{3}\,\partial^2\left[
 \frac{\xi\,(J_\chi-2\,H\,Q)}{J}\,\Pi^L + \frac{r}{a\,(r+1)}\left(
 B + \frac{1}{a}\,\Pi^0 + \frac{a}{r}\,\dot\Pi^L
 \right)
\right]
&=0\,.
\label{eq:Pcons}
\end{align}

In order to reduce the action to include only the dynamical modes $\Pi^L$ and $\delta\zeta$, we need to solve the above equations for $\Phi$, $B$ and $\Pi^0$. However, we note that the full solutions of $\Phi$, $B$ and $\Pi^0$ involve inverse Laplacians. This is manageable for the quadratic action: at linear order different scales do not interact, thus the dynamics can be described in terms of Fourier modes. However, since we aim to derive the interactions, we simplify the calculation by considering short scales. To this goal, we apply the gradient expansion, where spatial gradients are much larger than the Hubble and mass scales. 
A detailed calculation including the next-to-leading order terms is presented in Appendix \ref{app:NTL-gradient}. Here, we only show the leading order terms, which corresponds to the sub-horizon limit.\footnote{We stress that this is not the quasi-static limit. We give the time and spatial derivatives of the perturbations the same weight since eventually we will consider a scalar mode sourced by propagating waves.}
As a further simplification, we also consider a smooth matter for which $\rho+P \to 0$.\footnote{Even in the sub-horizon approximation, matter perturbations are of the order of $\partial^2 \Phi$, so the solutions for $\Phi$, $B$ and $\Pi^0$ would each contain a term $\propto \partial^{-2}\delta\zeta$. This term can only contribute to the interactions involving the matter perturbation, which is not the focus of the present paper. }

We solve for Eq.\eqref{eq:Bcons} for $B$ and replace it in \eqref{eq:Fcons}. We then impose the following:
\begin{equation}
\left\vert\frac{3\,m^2J\,\xi}{2\,(1+r)}\Phi\right\vert \ll \left\vert\frac{\partial^2\Phi}{a^2}\right\vert\,,
\qquad
\left\vert\frac{3\,m^2J\,\xi}{2\,r}\Pi^0\right\vert \ll \left\vert\frac{\partial^2\Pi^0}{a^2}\right\vert\,.
\label{eq:SHapprox1}
\end{equation}
With this approximation, Eq.\eqref{eq:Fcons} can be integrated twice and solved for $\Phi$. Finally, we move on to Eq.\eqref{eq:Pcons} and use the solutions for $B$ and $\Phi$. We then impose
\begin{equation}
\left\vert3\,\xi\,H^2\left(\frac{m^2J}{2\,r\,H^2}+\frac{\left(2\,Q-\frac{J_\chi}{H}\right)(1+r)}{J\,r}\right)\Pi^0\right\vert\ll \left\vert \frac{\partial^2\Pi^0}{a^2}\right\vert\,,
\quad
\left\vert\frac{3\,m^2J\,\xi}{2\,r}\Pi^L\right\vert \ll \left\vert\frac{\partial^2\Pi^L}{a^2}\right\vert\,,
\quad
\left\vert\frac{3\,m^2J\,\xi}{2\,r}\dot{\Pi}^L\right\vert \ll \left\vert\frac{\partial^2\dot{\Pi}^L}{a^2}\right\vert\,,
\label{eq:SHapprox2}
\end{equation}
which reduce the equation to an algebraic equation in $\Pi^0$. The solutions for the non-dynamical modes are thus: 
\begin{align}
B =& -\frac{a\,m^2J\,\xi}{2\,H}\,\Pi^L\,,\nonumber\\
\Phi = & \frac{a^2m^2J\,\xi}{2\,H}\left[\dot{\Pi}^L -\left(\frac{m^2J(r-1)}{2\,H}+\frac{(2\,H\,Q-J_\chi)\,r}{J}\right)\xi\,\Pi^L\right]\,,\nonumber\\
\Pi^0 = & \frac{a^2}{r}\,\left[-\dot{\Pi}^L +\left(\frac{m^2J\,r}{2\,H}+\frac{(2\,H\,Q-J_\chi)(r+1)}{J}\right)\xi\,\Pi^L\right]\,.
\label{eq:SHsolutions}
\end{align}
These relations are valid in the regime where the conditions \eqref{eq:SHapprox1}-\eqref{eq:SHapprox2}, which can be written more compactly as
\footnote{We note that our approximation corresponds to the interval $ \lambda \ll \lambda_{GMG} < \lambda_H$ where $\lambda_{GMG} \equiv J/ H$ is the characteristic length associated with the GMG theory. Because of this, notice that the dRGT limit $J\to0$ is not described by this approximation. See Ref.\cite{Kenna-Allison:2019tbu} for a discussion of this point. 
 }
\begin{equation}
\left\vert\frac{m^2J\,\xi}{r}\right\vert \ll k^2/a^2\,,\qquad
\left\vert\frac{3\,\xi\,H^2(1+r)}{Jr}\left(2\,Q-\frac{J_\chi}{H}\right)\right\vert \ll k^2/a^2\,,
\label{eq:SHapproximations}
\end{equation}
hold. The first condition is trivially satisfied in self-accelerating backgrounds with $m \sim H_0$ where one has $0 < J < 1$ \cite{Kenna-Allison:2019tbu}.\footnote{
Note that in self-accelerating backgrounds both $\xi$ and $r>1$ acquire typically $\mathcal{O}(1)$ values \cite{Kenna-Allison:2019tbu}.}.  For the second one, $J$ appears in the denominator, therefore the left hand side is not necessarily small. On the other hand, for perturbations sourced by LIGO-scale gravitational waves, the right hand side would be $(k/a)^2 \sim \omega^2_{\rm LIGO} \sim \Lambda_3^2 \sim 10^{40} H_0^2$, so it is possible to keep the second condition valid to a very good accuracy. Nevertheless, this condition needs to be verified on a case-by-case basis to ensure that the approximations hold. We show an explicit example later in Sec.\ref{sec:example}.

Now equipped with the sub-horizon solutions \eqref{eq:SHsolutions}, we use them in the Lagrangian \eqref{eq:Lagrangian} to obtain:
\begin{equation}
\delta^2S = \frac{1}{2}\int d^3x\,dt\,a^3\mathcal{K}_\Pi\,\left[\left(\dot\Pi^L\right)^2 - \frac{c_s^2}{a^2} \,\partial_i\Pi^L\partial^i\Pi^L+ \mathcal{O}(\Pi^2)\right]\,,
\end{equation}
where $\Pi^2$ terms are suppressed in the sub-horizon approximation and
\begin{align}
\mathcal{K}_\Pi &= \frac{3\,m^2\Mpl^2 a^4\xi^2H^2}{r}\left(2\,Q - \frac{J_\chi}{H}+\frac{m^2J^2}{2\,H^2}\right)\,,\nonumber\\
c_s^2 & = \frac{r}{3\,H^2\,\left(2\,Q - \frac{J_\chi}{H}+\frac{m^2J^2}{2\,H^2}\right)} \Bigg[
 \frac32\,m^2J\left(J+2\,Q(r-1)\xi\right)-J_{\chi\chi}r\,\xi+\frac{(J_\chi-2\,H\,Q)^2(r-1)\xi}{J}
 \nonumber\\
&\qquad\qquad\qquad\qquad\qquad\qquad\qquad+2\,H^2(3\,Q-G\,\xi) +H\left(-3\,J_\chi+2\,Q_\chi(r+1)\xi\right)
 \Bigg]\,.
 \label{eq:kinpi}
\end{align}

We can thus define the canonically normalised scalar graviton as
\begin{equation}
\hat\Pi \equiv \sqrt{\mathcal{K}_\Pi}\,\Pi^L = \frac{\sqrt{3}\,(\Mpl m\,\xi\,a^2H)}{\sqrt{r}}\,\sqrt{2\,Q-\frac{J_\chi}{H}+\frac{m^2J^2}{2\,H^2}}\,\,\Pi^L \,,
\end{equation}
with the associated sound speed given by $c_s$.

\section{Interactions involving tensor modes}

In this section, we first calculate the cubic order tensor-scalar-scalar and tensor-tensor-scalar interactions in generalised massive gravity. The latter terms provide the mechanism for the tensors to source the scalar mode, while the former terms provide the channel for gravitational decay and/or lead to instability.

In order to reduce the non-linear action down to interactions between independent dynamical modes, we need to integrate out the non-dynamical scalar perturbations $\delta g_{00}$, $\delta g_{0i}$ and $\delta \phi ^0$, i.e. $\Phi$, $B$ and $\Pi^0$. Although we focus on interactions that appear at third order in perturbations, the constraints are needed only at linear order (i.e. they extremise the {\it quadratic} action). This is due to the coupling between the tensor and scalar appearing at cubic order and above, thus the high order corrections to the scalar modes only start contributing at quartic order. 

\subsection{Tensor-Scalar-Scalar vertex}
We expand the action to cubic order while keeping terms linear on the tensor modes,
\begin{equation}
\delta^{3}S\ni\frac{\Mpl^2}{2}\int dt\,a^3 d^3x\,\mathcal{L}^{tss}\,,
\end{equation}
where, after some manipulation with boundary terms and assuming smooth matter, we obtain
\begin{align}
%
%
\mathcal{L}^{tss} = & 
%
- \frac{4\,H}{a}\,h_{ij}\partial^i\Phi\,\partial^jB
-\frac{1}{a}\,\dot{h}_{ij}\partial^i\Phi\,\partial^jB
%
+m^2\,\xi\,(J-Q\,\xi)\,h_{ij}\Phi\,\partial^i\partial^j\Pi^L
\nonumber\\
&
-\frac{1}{a^2}\,h_{ij}\partial^i\partial^jB\,\partial^2B
+\frac{1}{a^2}\,h_{ij}\partial^j\partial^kB\,\partial_k\partial^iB
+\frac{m^2\,\xi\,\left[-J\,(3+4\,r) + Q\,\xi\,(r+1)\right]}{2\,(r+1)^2}\, h_{ij}\,\partial^iB\,\partial^jB
\nonumber\\
&
%
+\frac{m^2r^2\,\xi\,\left[-J\,(2\,r+1)+Q\,\xi\,(r+1)\right]}{a\,(r+1)^2}\,h_{ij}\,\partial^iB\,\partial^j\Pi^0
%
-\frac{m^2\,a\,\xi\,\left[-J\,(2\,r+1)+Q\,\xi\,(r+1)\right]}{(r+1)^2}\,h_{ij}\,\partial^iB\,\partial^j\dot{\Pi}^L
\nonumber\\
&
%
+\frac{m^2r^2\xi\,\left[-J\,(2\,r+1)+Q\,\xi\,(r+1)\right]}{2\,a^2(r+1)^2}\,h_{ij}\partial^i\Pi^0\,\partial^j\Pi^0
%
+m^2\,r\,\xi^2\left[J_\chi-Q_\chi\xi+H\,(-2\,Q+G\,\xi)\right]\,h_{ij}\Pi^0\partial^i\partial^j\Pi^L
\nonumber\\
&
-m^2Q\,r\,\xi^2 \dot{h}_{ij}\Pi^0\partial^i\partial^j\Pi^L
+\frac{m^2r\,\xi\,\left[-J\,r+Q\,\xi\,(r+1)\right]}{(r+1)^2}\,h_{ij}\partial^i\Pi^0\partial^j\dot{\Pi}^L
\nonumber\\
&
%
+\frac{m^2a^2\xi\,\left[J+Q\,\xi\,(r+1)\right]}{2\,(r+1)^2}\,h_{ij}\partial^i\dot{\Pi}^L\partial^j\dot{\Pi}^L
+m^2\xi^2\left[Q+G\,\xi\,(r-1)\right]\left(h_{ij}\partial^i\partial^j\Pi^L\partial^2\Pi^L-h_{ij}\partial^j\partial^k\Pi^L\partial_k\partial^i\Pi^L\right)\,.
\label{eq:Lag_tss}
\end{align}
Once the linear solutions \eqref{eq:SHsolutions} for the non-dynamical modes $B$, $\Phi$ and $\Pi^0$ are substituted, the cubic tensor-scalar-scalar interactions reduce to
\begin{align}
\delta^{3}S\ni\frac{1}{\sqrt{2}}\int a^3 d^4x 
\left[
\frac{1}{\Lambda_3^3}\left(C_{tt}\partial_t^2 + C_{\partial\partial}\,\frac{\partial^2}{a^2}\right) \hat{h}_{ij} 
+ \frac{C_{t}}{\Lambda_2^2} \,\dot{\hat{h}}_{ij} + 
\mathcal{O}\left(\hat{h}_{ij}\right)
\right]\,\frac{\partial^i\hat{\Pi}\,\partial^j\hat{\Pi}}{a^2}
\label{eq:cubicTSS}
\end{align}
where we have introduced the canonically normalised perturbations through:
\begin{align}
h_{ij} =& \frac{\sqrt{2}}{\Mpl}\hat{h}_{ij}\,,\nonumber\\
\Pi^L = & \frac{\sqrt{r}}{\sqrt{3}\,(\Mpl m\,\xi\,a^2H)}\,\left(2\,Q-\frac{J_\chi}{H}+\frac{m^2J^2}{2\,H^2}\right)^{-1/2}\hat{\Pi}\,.
\end{align}
In Eq.\eqref{eq:cubicTSS}, we explicitly present the 3- and 4-derivative interactions, which only get contribution from leading order terms in gradient expansion (sub-horizon approximation).\footnote{\label{footy}Next-to-leading order corrections in $\Phi$ and $\Pi^0$ reduce the order of derivative by at least 1, whereas $B$ reduces it by 2 (see Appendix \ref{app:NTL-gradient}). From Eq.\eqref{eq:Lag_tss}, we see that the next-to-leading order contributions do not affect the 3-derivative terms in \eqref{eq:cubicTSS}.}

The coefficients of the above terms are given by:
\begin{align}
 C_{t} \equiv &
 \frac{Q\,r\,(r+1)\,\xi}{3\,\hh\,J}
 -\frac{\left[3\,J^2+4(J\,Q-\hh\,Q_c)(r+1)\xi + 4\,\hh^2(-3\,Q+2\,G\,\xi)\right]\,r}{12\,\hh\,\left(J^2-2\,\hh\,J_c+4\,\hh^2\,Q\right)}\,,
 \nonumber\\
 C_{tt} \equiv  & \nonumber -\frac{(J^2-4\,\hh^2Q)\,r}{12\,\hh^2\left(J^2 -2 \hh\,J_c+4\,\hh^2\,Q\right)}\,,\\
 C_{\partial\partial} \equiv & \frac{\left(J^2 -4\,\hh^2Q -4\,\hh^2 G(r-1)\xi\right)\,r}{12\,\hh^2(J^2-2\,\hh\,J_c+4\,\hh^2Q)}\,,
 \label{eq:coeffs}
\end{align}
with $\Lambda_n \equiv (\Mpl m^{n-1})^{1/n}$ and we defined the following dimensionless quantities 
\begin{equation}
\hh \equiv \frac{H}{m}
\,,\;\;
J_{c} \equiv  \frac{J_\chi}{m}
\,,\;\;
Q_{c} \equiv  \frac{Q_\chi}{m}
\,.
\end{equation}

\subsection{Tensor-Tensor-Scalar interactions}
\label{sec:TTS-interactions}
To be able to perform the calculation analogous to the scalar-tensor case \cite{Creminelli:2019kjy}, we also compute the cubic interactions that involve two tensor modes. We present the tensor-tensor-scalar Lagrangian before integrating out the non-dynamical degrees in Appendix \ref{app:TTS}. After the reduction, then switching to canonically normalised fields, we find
\begin{align}
\delta^{3}S\ni\int a^3 d^4x \Bigg[&
\frac{\mathcal{A}_3 }{\Lambda_2^2}\,\hat{\Pi}\,\dot{\hat{h}}^{ij}\,\left(
- \ddot{\hat{h}}_{ij} + \frac{1}{a^2}\,\partial^2\hat{h}_{ij}
\right)
\nonumber\\
&+\frac{1}{\Mpl}\,\hat\Pi\,\left[
\mathcal{A}_{2a}\,\dot{\hat{h}}_{ij}\dot{\hat{h}}^{ij}
+\mathcal{A}_{2b}\,\ddot{\hat{h}}_{ij}\hat{h}^{ij}
+\frac{\mathcal{A}_{2c}}{a^2}\,
\left(\partial_i\hat{h}_{jk}\partial^i\hat{h}^{jk}-\partial_i\hat{h}_{jk}\partial^j\hat{h}^{ik}\right)
+\frac{\mathcal{A}_{2d}}{a^2}\,\partial^2\hat{h}_{ij}\hat{h}^{ij}
\right]
\nonumber\\
&+\mathcal{A}_1 \,\left\{\frac{m\,\partial \hat{\Pi}\,\hat{h}\,\hat{h}}{\Mpl}\right\}
+\mathcal{A}_0 \,\left\{\frac{m^2\, \hat{\Pi}\,\hat{h}\,\hat{h}}{\Mpl}\right\}
\Bigg]\,,
\label{eq:TTSterms}
\end{align}
where we defined
\begin{align}
\mathcal{A}_3 &= -\frac{J\,\sqrt{r}}{2\,\sqrt{6}\,\hh\,\sqrt{J^2-2\,\hh\,J_c+4\,\hh^2Q}}\,,\nonumber\\
\mathcal{A}_{2a} &= \frac{\sqrt{r}\,\left[2\,J+Q\,(r+1)\,\xi\right]}{2\,\sqrt{6}\,\sqrt{J^2-2\,\hh\,J_c+4\,\hh^2Q}}\,,\nonumber\\
\mathcal{A}_{2b} &= \frac{\sqrt{r}\,Q\,\xi}{\sqrt{6}\,\sqrt{J^2-2\,\hh\,J_c+4\,\hh^2Q}}\,,\nonumber\\
\mathcal{A}_{2c} &= \frac{\sqrt{r}\,\left[J +Q\,(r-3)\,\xi-2\,G\,(r-1)\,\xi^2\right]}{2\,\sqrt{6}\,\sqrt{J^2-2\,\hh\,J_c+4\,\hh^2Q}}\,,\nonumber\\
\mathcal{A}_{2d} &= -\frac{\sqrt{r}\,\left[Q+G\,(r-1)\,\xi\right]}{\sqrt{6}\,\sqrt{J^2-2\,\hh\,J_c+4\,\hh^2Q}}\,,\end{align}
and we did not explicitly present terms lower than 2-derivative orders in \eqref{eq:TTSterms} since they are further suppressed. 
Although the 2-derivative terms are also suppressed compared to the 3-derivative ones, they are nevertheless accurate: applying the argument in footnote \ref{footy} to the tensor-tensor-scalar action \eqref{eq:TTSterms-pre}, we conclude that the 2-derivative terms given in \eqref{eq:TTSterms} do not get any contribution from next-to-leading order corrections in gradient expansion.

Analogously in Beyond Horndeski, the 3-derivative term $\dot{\Pi}\,\dot{h}^2$ is responsible for sourcing the scalar mode, while the 2-derivative term $\Pi\,\dot{h}^2$  has the same effect in cubic Galileon theory \cite{Creminelli:2019kjy}.

\subsection{Higher order interactions with a single tensor}\label{sec:high_order}

In massive gravity, the vertices with a single tensor do not stop at cubic order. In this section, we show the evidence of contributions of the form $h (\partial\partial\Pi)^{n-1}$.  There is however one setback: the solutions \eqref{eq:SHsolutions} to the constraint equations were calculated using linear equations of motion. In order to compute interactions with $n>3$, we need to find high order corrections to the constraint equations. For instance, the $h_{ij}\,\partial^i \Pi^0 \partial^i \Pi^0$ term in Eq.\eqref{eq:Lag_tss} contributes not only to the $h_{ij}\,\partial^i\dot\Pi \partial^j \dot\Pi$ interaction, but also to other terms $h\,(\partial_i\dot\Pi)^{n-1}$ terms with $n>3$ through the non-linear corrections.

With our current knowledge however, by disregarding the non-linear corrections in the non-dynamical degrees of freedom, we can still look at an isolated piece of the action which directly contributes to $h (\partial\partial\Pi)^{n-1}$ interactions, where $\partial$ without indices represents either a spatial or time derivative.
The presence of these terms in this simplified case, as well as the order of their coefficients, hints at their existence in the complete calculation.\footnote{We remark that one should not be tempted to determine the non-linear dynamics of $\Pi$ using only the partial Largrangians presented here. Since the necessary counter-terms are missing, one might end up with a derivative order spuriously higher than two.}

At $n=4$, the terms that contain 6 derivatives are formally of type $h (\partial\partial \Pi) \,(\partial\dot\Pi)^2$ which we find to be
\begin{equation}
S^{(4)} = \int d^4x a^3 \,\mathcal{L}^{(4)}\,,
\end{equation}
with
\begin{align}
\mathcal{L}^{(4)} \ni
\frac{1}{\Lambda_3^6}\,\frac{r^{3/2}}{3\,\sqrt{3}\,(r+1)\,\xi^2\,a^4\,\mathcal{N}^{3/2}}\,h_{ij}
\Bigg[
&
(-Q\,\xi+G\,\xi^2)\partial^i\partial^j\Pi\,\partial_k\dot\Pi\,\partial^k\dot\Pi 
+\left(\frac{Q\,\xi}{r+1}+G\,\xi^2\right)\,\partial^i\dot\Pi\,\partial_j\dot\Pi\,\partial^2\Pi
\nonumber\\
&+ 
\left(\frac{J(r-1)}{(r+1)^2} + \frac{Q(r-1)\xi}{r+1}-2\,G\,\xi^2\right)\,\partial^i\dot\Pi\,\partial^j\partial^k\Pi\,\partial_k\dot\Pi
\Bigg]\,,
\label{eq:quartic}
\end{align}
where $\mathcal{N} \equiv J^2 - 2\,h\,J_c+4\,h^2 Q$ and for clarity of notation, we suppressed the hat in the canonically normalised fields. 

At quintic order, 8-derivative terms are formally of types $h (\partial\partial \Pi)^2 \,(\partial\dot\Pi)^2$ and $h (\partial\dot\Pi)^4$ which are:
\begin{equation}
S^{(5)} = \int d^4x a^3 \,\mathcal{L}^{(5)}\,,
\end{equation}
with
\begin{align}
\mathcal{L}^{(5)} \ni
\frac{1}{\Lambda_3^9}\,\frac{\sqrt{2}\,r^2}{9\,(r+1)^2 \xi^3 a^6 \mathcal{N}^2}\,h_{ij}
\Bigg[
&
-G\,\xi^2(r+2)\,\partial^i\partial^j\Pi\,\partial^2\Pi\,\partial_k\dot\Pi\,\partial^k\dot\Pi
+[Q\,\xi+G\,\xi^2(r+1)]\,\partial^i\partial^j\Pi\,\partial_k\dot\Pi\partial^k\partial^l\Pi\,\partial^l\dot\Pi
\nonumber\\
&
+G\,\xi^2\,(r+2)\,\partial^i\partial^k\Pi\,\partial_k\partial^j\Pi\,\partial_l\dot\Pi\,\partial^l\dot\Pi
\nonumber\\
&
+\left(
\frac{J\,(r^2+1)}{2(r+1)^2}+\frac{Q\,\xi\,(r^2+1)}{2\,(r+1)}-G\,\xi^2(r+1)\right)
\partial^i\partial^k\Pi\,\partial^j\partial^l\Pi\,\partial_k\dot\Pi\partial_l\dot\Pi
\nonumber\\
&
+\left(
\frac{J\,(1-r(4+r))}{2(r+1)^2}-\frac{Q\,\xi\,(-1+r(4+r))}{2\,(r+1)}-G\,\xi^2(r+1)\right)
\partial^i\dot\Pi\, \partial^j\partial^k\Pi\,\partial_k\partial^l\Pi\,\partial_l\dot\Pi
\nonumber\\
&+\left(\frac{Q\,\xi(r-1)}{r+1}+G\,\xi^2(r+1)\right)\partial^i\dot\Pi \partial^j\partial^k\Pi\, \partial_k\dot\Pi\,\partial^2\Pi
+\frac{a^2[3\,J+Q(r+1)\xi]}{4\,(r+1)^2}\,\partial^i\dot\Pi\,\partial^j\dot\Pi \partial^k\dot\Pi\partial_k\dot\Pi
\Bigg]
\label{eq:quintic}
\end{align}
This straightforward calculation indicates that high derivative interactions of the type
\begin{equation}
\mathcal{L}^{(n)} \ni \frac{1}{\Lambda_3^{3(n-2)}}\,h \,(\partial\partial\Pi)^{n-1}\,,
\end{equation}
do exist with an associated strong-coupling scale at a low $\Lambda_3$, where we remind the reader that $\partial$ here denotes either a spatial or time derivative.

\section{Interactions in the presence of a propagating tensor wave}
\subsection{Sourcing the scalar mode}
We now discuss the implications of a LIGO-scale gravitational waves on the scalar mode. Initially, $\hat\Pi$ would be in its vacuum state, but once the gravitational waves are emitted, it will be sourced by the tensor-tensor-scalar interactions. In this context, the relevant Lagrangian is
\begin{align}
 \mathcal{L} =& \frac12\,\dot{\Pi}^2-\frac{1}{2}\,c_s^2\,\partial_i\Pi\,\partial^i\Pi + \frac{\mathcal{A}_3}{\Lambda_2^2}\,\Pi\,\dot{h}_{ij}\,\left(-\partial_t^2+ \frac{\partial^2}{a^2}\right)h^{ij}
 \nonumber\\
&+\frac{1}{\Mpl}\,\Pi\,\left[
\mathcal{A}_{2a}\,\dot{h}_{ij}\dot{h}^{ij}
+\mathcal{A}_{2b}\,\ddot{h}_{ij}h^{ij}
+\frac{\mathcal{A}_{2c}}{a^2}\,
\left(\partial_ih_{jk}\partial^ih^{jk}-\partial_ih_{jk}\partial^jh^{ik}\right)
+\frac{\mathcal{A}_{2d}}{a^2}\,\partial^2h_{ij}h^{ij}
\right]
\,,
\label{eq:sourcing-pi}
\end{align}
where once again, we omit the over-hat in the canonically normalised fields. 
Calculating the equation of motion for $\Pi$ and using the plane-wave solution for $h_{ij}$ propagating in a Minkowski background along the $\hat{z}$ direction with linear $+$ polarisation, given by \cite{Creminelli:2019nok}
\begin{equation}
h_{ij} = \Mpl h_0^+\,\sin[\omega(t-z)]\,\epsilon^+_{ij}\,,
\label{eq:hij}
\end{equation}
we find that the leading 3-derivative interaction terms $\propto \mathcal{A}_3$ cancel. We therefore need to go to next-to-leading order. Although the 2-derivative terms are still valid at leading order in gradient expansion, the plane wave solution \eqref{eq:hij} is not: we need to relax the assumption that the gravitational waves propagate in a flat spacetime. Including the Hubble friction term in the equation of motion of tensor modes leads to a dependence on the scale factor as $a^{-3/2}$ in Eq.\eqref{eq:hij}. Thus, the term $\dot{h}_{ij}(-\partial_t^2 + \partial^2/a^2) h_{ij}$ will effectively give a contribution of $ 3\,H\,\dot{h}_{ij}\dot{h}^{ij}$ at next-to-leading order. In other words, the coefficient of the corresponding 2-derivative term will be altered as:
\begin{equation}
\mathcal{A}_{2a} \to \tilde{\mathcal{A}}_{2a} =  \mathcal{A}_{2a} + \frac{3\,H}{m}\,\mathcal{A}_3 = \frac{\sqrt{r}\,\left[-J+Q\,(r+1)\,\xi\right]}{2\,\sqrt{6}\,\sqrt{J^2-2\,h\,J_c+4\,h^2Q}}\,.
\end{equation}
With this correction, we can now calculate the equation of motion for $\Pi$ as:
\begin{equation}
\ddot{\Pi} - c_s^2 \partial^2{\Pi} = \frac{\omega^2\Mpl(h_0^+)^2}{2} \Big(\mathcal{A}_{2-}
+\mathcal{A}_{2+}\,\cos\left[2\,\omega\,(t-z)\right]\Big)\,,
\end{equation}
where
\begin{align}
 \mathcal{A}_{2-} \equiv\tilde{\mathcal{A}}_{2a}+\mathcal{A}_{2c} - \mathcal{A}_{2b}-\mathcal{A}_{2d} =&
  \frac{\sqrt{r}\,Q\,(r-1)\,\xi}{\sqrt{6}\,\sqrt{J^2-2\,h\,J_c+4\,h^2Q}}\,,
  \nonumber\\
  \mathcal{A}_{2+} \equiv\tilde{\mathcal{A}}_{2a}+\mathcal{A}_{2c} + \mathcal{A}_{2b}+\mathcal{A}_{2d} =&
  \frac{\sqrt{r}\,(Q-2\,G\,\xi)\,(r-1)\,\xi}{\sqrt{6}\,\sqrt{J^2-2\,h\,J_c+4\,h^2Q}}\,,
\end{align}
As we see above, the source term has two components: a constant part and a part that oscillates (propagates) with frequency $2\,\omega$. Thus the solution will also have two components: a growing part that goes as $t^2$ and an oscillating part with the same frequency and a constant amplitude. We can find a particular solution by assuming $\mathcal{A}_i$ are constant in time and $\Pi$ depends only on $t-z$ :
\begin{equation}
\Pi = \frac{\omega^2\Mpl(h_0^+)^2}{4(1-c_s^2)} \mathcal{A}_{2-} \,(t-z)^2 + \frac{\Mpl(h_0^+)^2}{8\,(c_s^2-1)}\mathcal{A}_{2+}\,\cos\left[2\,\omega\,(t-z)\right]\,.
\label{eq:Pisolution}
\end{equation}
For null coordinates we have  $\omega(t-z)\sim \mathcal{O}(1)$, thus $\Pi$ approximately propagates with twice the frequency of the tensor wave, with an amplitude of $\Mpl (h_0^+)^2\mathcal{A}_{2+} /[8\,(c_s^2-1)]$.

This is reminiscent of the result of Ref.~\cite{Creminelli:2019kjy} for cubic Galileons, with $\mathcal{A}_{2-}=\mathcal{A}_{2+}$. However, there are a couple of differences. First, Ref.~\cite{Creminelli:2019kjy} never needed to compute the solution explicitly, but used the equation of motion for sourced $\Pi$ to show that the deviations around this solution are unstable. Secondly, they confirmed that the terms quadratic in $\Pi$ in the equation of motion do not contribute to the sourced solution. For GMG, these terms were computed under the quasi-static approximation in \cite{Gumrukcuoglu:2021gua}. However, we cannot ignore time derivatives of $\Pi$ in the present case. Although these terms can be computed perturbatively, we expect that there are infinite self-interaction terms of $\Pi$ and that the perturbative expansion breaks down. Therefore, without an approach that considers contributions from all orders in perturbation theory, we are unable to compute these purely $\Pi$ non-linear terms in this context.

\subsection{Instability induced by gravitational waves}
In the presence of the gravitational waves, the interactions \eqref{eq:cubicTSS} modify the effective sound-speed of $\Pi$. To determine the extent of this modification, we will compare each tensor-scalar-scalar interaction with the free Lagrangian
\begin{equation}
 \mathcal{L}_\Pi^{(2)} = \frac12\, \dot\Pi^2 - \frac{c_s^2}{2}\,\partial_i\Pi\,\partial^i\Pi \sim \frac{1-c_s^2}{2}\partial_i\Pi\partial^i\Pi\,.
\end{equation}
However, we first discuss the correction to the 2-derivative interaction from the effect of tensor modes propagating in an expanding background. In this case, the term $ \ddot{h}_{ij}\partial^i\Pi\partial^j\Pi$ in Eq.\eqref{eq:cubicTSS} will also give a contribution $-3\,H\,\dot{h}_{ij}\partial^i\Pi\partial^j\Pi$. This can be absorbed in a new definition for the coefficient $C_t$ as
\begin{equation}
 C_t\to \tilde{C}_t = C_t - \frac{3\,H}{m}\,C_{tt} =
 \frac{Q\,r\,(r+1)\,\xi}{3\,\hh\,J}
 -\frac{\left[(J\,Q-\hh\,Q_c)(r+1) + 2\,\hh^2G\right]\,r\,\xi}{3\,\hh\,\left(J^2-2\,\hh\,J_c+4\,\hh^2\,Q\right)}\,.
\end{equation}
Thus, we compare the free Lagrangian with the amplitudes of the tensor-scalar-scalar interactions
\begin{align}
\frac{\mathcal{L}_{\partial^4 h\Pi\Pi}}{ \mathcal{L}_\Pi^{(2)} }
&\sim \left(\frac{\omega}{m}\right)^2h_0^+ \frac{C_{tt}+C_{\partial\partial}}{(1-c_s^2)} \sim 
10^{20}\, \frac{C_{tt}+C_{\partial\partial}}{(1-c_s^2)} 
\,,
\label{eq:gradient-L3}
\\
\frac{\mathcal{L}_{\partial^3 h\Pi\Pi}}{ \mathcal{L}_\Pi^{(2)} } &\sim 
\frac{\omega}{m}\,h_0^+ \frac{\tilde{C}_t}{(1-c_s^2)} 
\sim
\frac{\tilde{C}_{t}}{(1-c_s^2)}\,,
\label{eq:gradient-L2}
\end{align}
where we specified to a LIGO-scale gravitational wave with $\omega/H_0 \sim 10^{20}$ and $h_0^+ \sim 10^{-20}$ and assumed $m\sim H_0$. In this case, the cubic interaction with the external gravitational waves dominates the gradient energy of the $\Pi$ perturbation. If these interactions are truncated at this order, this is a drastic result: since the gradient energy is proportional to $h_{ij}$, which is an oscillating function, there are times where the energy of the $\Pi$ field is unbounded from below, signaling an instability akin to the one presented in Ref.\cite{Creminelli:2019kjy}. 

In order to avoid the instability, one option would be to tune away the coefficients. In the case of the 4-derivative term in \eqref{eq:gradient-L3}, the coefficient is
\begin{equation}
C_{tt} + C_{\partial\partial} = -\frac{G\,(r-1)\,r\,\xi}{3(J^2-2\,\hh^2J_c+4\,\hh^2Q)}\,,
\label{eq:coefcombined}
\end{equation}
i.e. proportional to the function $G$. From its definition \eqref{eq:defLJQG}, $G$ can be tuned to zero by choosing the mass functions to satisfy $\alpha_3 +\alpha_4=0$. However, this choice might not be compatible with the linear stability considerations of a cosmological background. For instance, in the minimal model studied in Ref.\cite{Kenna-Allison:2019tbu}, where only $\alpha_2$ varies with a linear dependence on $\Phi^a\Phi_a$, this option compromises perturbative stability of the background. 

Although the 3-derivative interaction is not enhanced by the tensor frequency, it is not suppressed either. Therefore even if we can tune away the 4-derivative term, the 3-derivative term might become dominant. In this case, the coefficient $\tilde{C}_{t}$ cannot be tuned to vanish by a simple choice of mass functions since it has a non-trivial dependence on background quantities. However, it may be possible to find a region in the theory space where this combination stays small. 

A striking difference with the scalar-tensor examples is that in massive gravity, tuning away the interaction terms does not remove the source terms.  In particular, even after setting $C_{tt}+C_{\partial\partial} = \tilde{C}_t=0$, the combination of source term coefficients $\mathcal{A}_{2+}$ stays proportional to the untuned $Q$ function. 
This is where any attempt to construct an analogy with the scalar-tensor theories breaks down. For instance, in Beyond-Horndeski theory, the analogues of the $C_{tt}$ term and $\mathcal{A}_3$ have proportional coefficients, while in our case, not only this is not satisfied, we also find that the 3-derivative source terms have zero effect at leading order in the gradient expansion. Instead, at next order we have $\mathcal{A}_{2\pm}$ neither of which vanish when $C_{tt}+C_{\partial\partial}$ is set to zero. We stress once more that massive gravity is a distinct $\Lambda_3$ theory that simply cannot be interpreted in the scalar-tensor framework. 

We now turn to the magnitudes of the high order interactions that involve a single tensor mode, discussed in Sec.\ref{sec:high_order}. As an estimate for $\Pi$ in the presence of a LIGO-scale gravitational wave, we take the coefficient of the oscillating part in Eq.\eqref{eq:Pisolution}, i.e. $\Pi \sim \Mpl (h_0^+)^2 \mathcal{A}_{2+}$, and any derivative acting on $\Pi$ brings a factor of $\omega$. We consider the potentially dangerous interactions of the form 
\begin{equation}
 \mathcal{L}^{(n)} = \frac{C_n}{\Lambda_3^{3(n-2)}}\,h\,(\partial\partial\Pi)^{n-1}\,,
 \label{eq:higher}
\end{equation}
where $C_n$ represents the combined coefficients of same order terms. The  magnitude of these interactions are estimated as
\begin{equation}
\mathcal{L}^{(n)} \sim 
\Mpl^2m^2\,C_n\,\mathcal{A}_{2+}^{n-1}\, \left(\frac{\omega}{m}\right)^{2(n-1)}\,(h_0^+)^{2n-1}\,
\end{equation}
We note that as the order increases by 1, a factor of $(\partial\partial\Pi)/\Lambda_3^3$ is introduced, evaluating to
\begin{equation}
\frac{\partial\partial\Pi}{\Lambda_3^3} 
\sim \mathcal{A}_{2+}\,\left(\frac{\omega}{m}\right)^2(h_0^+)^2 \sim\mathcal{O}(1)\,\mathcal{A}_{2+}\,,
\label{eq:DDPI}
\end{equation}
where we used $\omega/m \sim 10^{20}$ and $h_0^+\sim 10^{-20}$. In other words, as the order increases there is no enhancement or suppression due to the frequency of the modes. To put it in another way:
\begin{equation}
 \frac{\mathcal{L}^{(n+1)}}{\mathcal{L}^{(n)}} \sim \left(\frac{\omega\,h_0^+}{m}\right)^2\,\frac{C_{n+1}\,\mathcal{A}_{2+}}{C_{n}} \sim \frac{C_{n+1}\,\mathcal{A}_{2+}}{C_{n}}\,.
\label{eq:strong}
\end{equation}
Thus, we see that for $\mathcal{O}(1)$ coefficients $C_n$ and $\mathcal{A}_{2+}$, the perturbative expansion breaks down.  

To summarise, the linear solution \eqref{eq:Pisolution}, generated at the onset of gravitational wave propagation, can still cause the non-linear self-interactions of $\Pi$ to become dominant, even before the instabilities are taken into account. Once the instability starts to act, we do not have any way of determining its fate since we have already lost perturbative control.

\subsection{An example: the minimal cosmological model}
\label{sec:example}
Up to now, we presented our results without specifying the theory in the GMG framework. In order to get a sense of the order of magnitude of the various coefficients that appear in the discussion above, we consider a specific background evolution in the minimal cosmological model studied in detail in Refs.\cite{Kenna-Allison:2019tbu, Kenna-Allison:2020egn}. This model is defined by the choice:
\begin{equation}
\alpha_0 = \alpha_1 = \alpha_3 =0\,,\qquad \alpha_4 = 0.8\,,\qquad \alpha_2 = 1+ 10^{-4} \phi^a\phi_a\,.
\end{equation}
Note that in our calculation so far, we neglected the matter and used the zero-curvature limit of Ref.\cite{deRham:2014gla}. These simplifications are valid at the scales relevant for our problem. However, a consistent evolution of the background in generalised massive gravity that is different than dRGT theory requires some matter field and a non-zero curvature. For this reason, we reintroduce matter and curvature with density parameters $\Omega_{m}=0.3$ and $\Omega_{K,0} = 3\times 10^{-3}$, respectively for the numerical evolution of the background.

\begin{figure}[h!]
	\centering
	\includegraphics[]{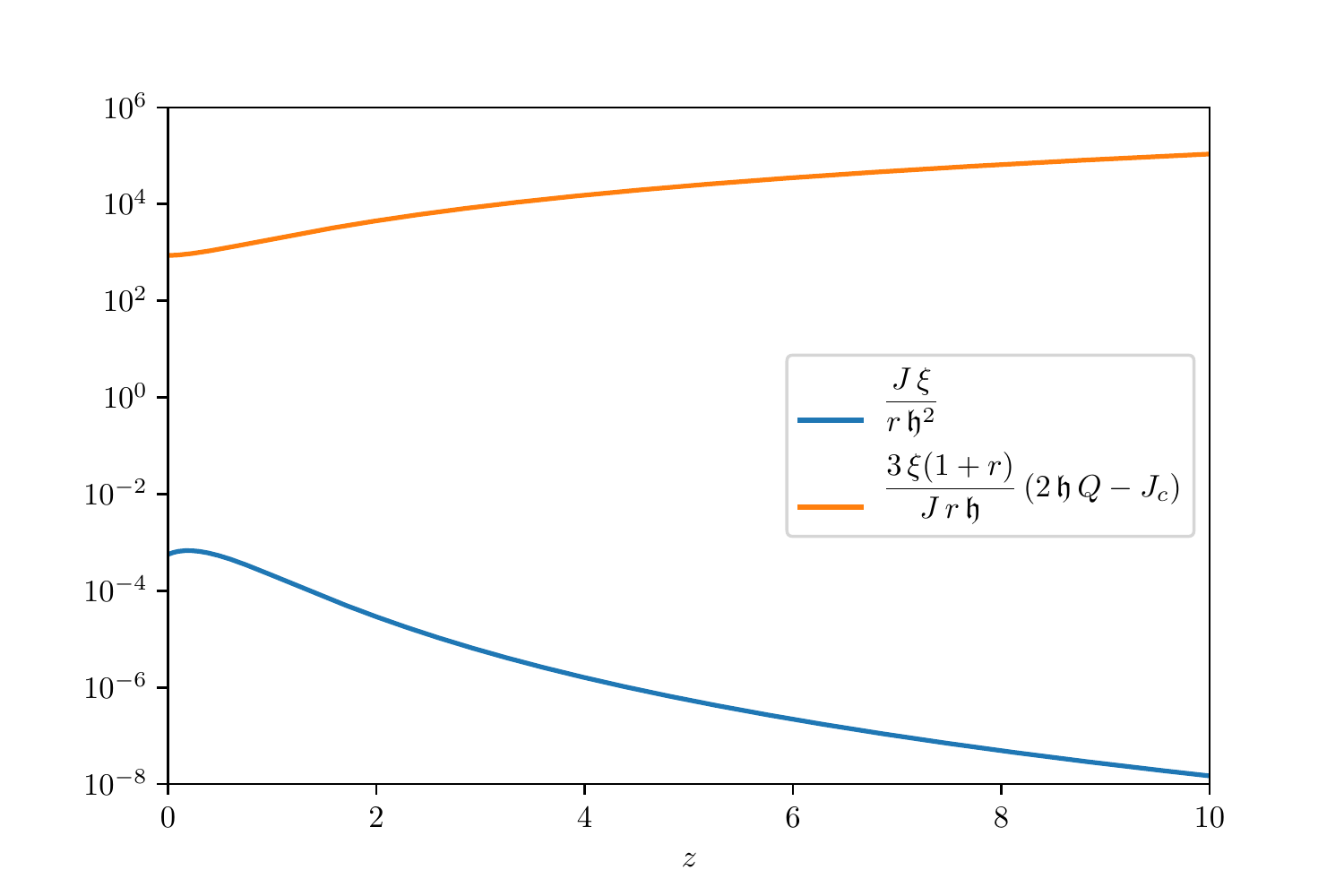}
	\caption{The evolution of the left hand sides of the two conditions \eqref{eq:SHapproximations} which are compared to $(k/aH)^2$. In this example, we adopt the parameters of Ref.\cite{Kenna-Allison:2020egn}.}
	\label{fig:approximation2}
\end{figure}

We first start by checking the conditions \eqref{eq:SHapproximations}, i.e. the validity of the gradient expansion. In Fig.\ref{fig:approximation2}, we show the left hand sides of these conditions where the right hand side is scaled to $(k/aH)^2$. 
We see that the first condition is satisfied as expected, while the left hand side of the second condition is relatively large, with a value $\sim10^3$ at $z=0$. This is still far below the scale for LIGO gravitational waves for which $k/aH_0\sim \Lambda_3/H_0 \sim 10^{20}$. Therefore, for this example, the subhorizon approximation is satisfied to a good accuracy.

\begin{figure}[h!]
	\centering
	\includegraphics[]{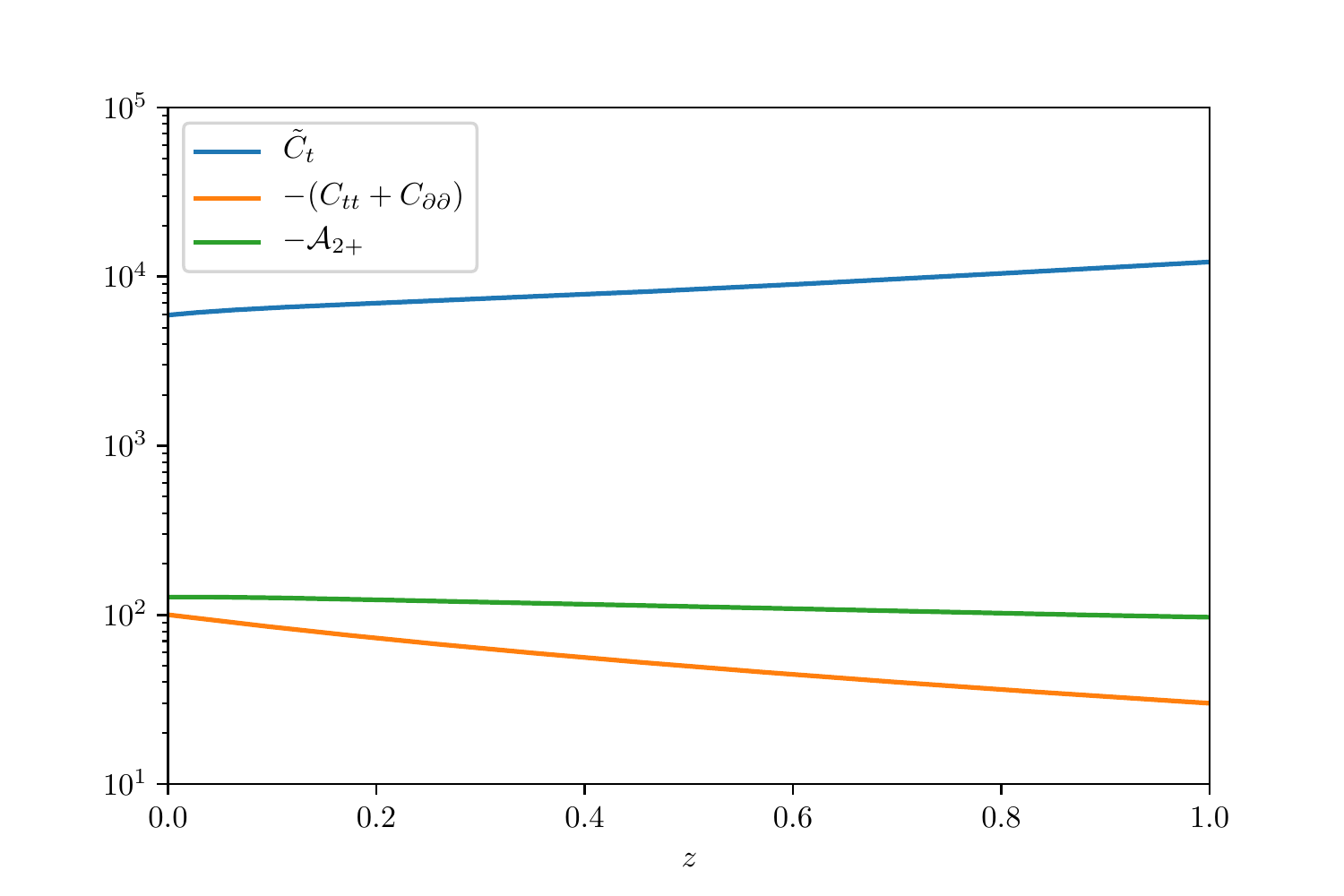}
	\caption{The evolution of the three relevant coefficients. All of them are of order $\mathcal{O}(10)$ or higher.}
	\label{fig:coeffs}
\end{figure}

Next, we consider the evolution of the three coefficients relevant for this analysis. These are:  $C_{tt}+C_{\partial\partial}$, the coefficient of the 4-derivative interaction term; $\tilde{C}_t$, the coefficient of the 3-derivative term; and $\mathcal{A}_{2+}$ which appears in the amplitude of the sourced $\Pi$ perturbation. The evolution of these coefficients between $z \in [0,1]$ is demonstrated in Fig. \ref{fig:coeffs}.
The large value of $|C_{tt}+C_{\partial\partial}| \sim O(10^2)$ implies that the 4-derivative interaction term gives the dominant contribution to the tensor-scalar-scalar interactions at cubic order as this term is enhanced by $10^{20}$ in Eq.~(\ref{eq:gradient-L3}). This would lead to a strong gradient instability. On the other hand, the large value of $|\mathcal{A}_{2+}| \sim O(10^2)$ means that the higher order interactions are not suppressed as shown in Eq.~(\ref{eq:strong}) unless there are unexpected cancellations for $C_n$. This makes it difficult to predict the fate of this instability.

\begin{figure}[h!]
	\centering
	\includegraphics[]{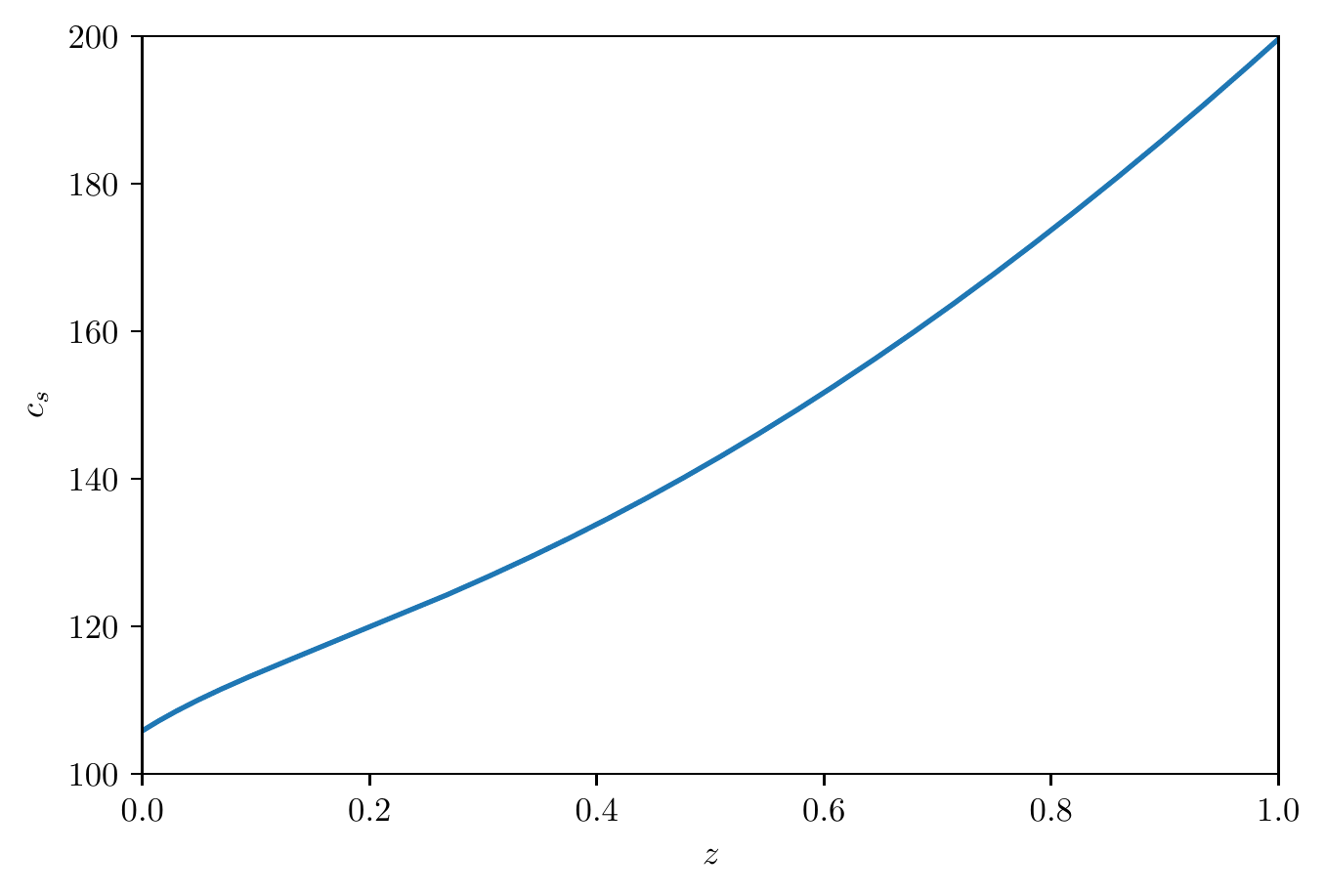}
	\caption{The evolution of the sound speed \eqref{eq:kinpi} for the scalar graviton $\Pi$. For this example, we used the same parameters as in the remainder of the plots. It should be noted that the plotted sound speed corresponds to the case where matter is simply a bulk cosmological constant, even though the background evolution requires some amount of matter and curvature for consistency.}
	\label{fig:cs2}
\end{figure}

It is also instructive to look at the sound speed of the scalar mode \eqref{eq:kinpi}. In Fig.\ref{fig:cs2}, we present the evolution of the scalar sound-speed. Notably, the sound speed at $z=0$ is $c_s\sim100$, i.e. it propagates at a speed $100$ times the speed of light. This is not surprising: since the background is very close to dRGT, the value of the sound speed carries the imprint of the strong coupling issue encountered in the dRGT limit (and the early asymptotics of GMG). The only way to have sub-luminal sound speed is to divert considerably from the self-accelerating solutions. This is an important point since the gravitational wave decays described in Refs.~\cite{Creminelli:2018xsv, Creminelli:2019nok} will be kinematically forbidden for a scalar with a super-luminal sound speed. Note however that the gradient instability is not sensitive to the value of $c_s$. For finite sound speeds, the correction to the gradient term is an oscillating function and can dominate for sufficiently large value of gravitational wave frequency.

\section{Discussion}
In this paper, we studied interactions between tensor and scalar perturbations in $\Lambda_3$ models of dark energy. Constructed within frameworks such as scalar-tensor or massive gravity theory classes, these models have non-linear scalar interactions suppressed by $\Lambda_3^3 = (m^2 \Mpl)$ where $m \sim H_0$ is a model parameter. These interactions are responsible for the Vainshtein mechanism that decouples the scalar perturbations from a matter source. The scalar-tensor interactions are well understood in scalar-tensor theories and it was shown to provide strong constraints on the model parameters \cite{Creminelli:2018xsv, Creminelli:2019nok, Creminelli:2019kjy}. In particular, the tensor-scalar-scalar interaction involving two derivatives of the tensor perturbation is enhanced by the ratio $\omega/m$ where $\omega$ is the frequency of tensor perturbations. For LIGO gravitational waves, this ratio is given by $10^{20}$ for $m \sim H_0$, giving $\mathcal{O}(10^{-20})$ constraints on model parameters. To investigate the generality of this constraint, we considered generalised massive gravity (GMG) models as an example of $\Lambda_3$ models that is distinct from scalar-tensor theories. 

We found structures of tensor-scalar-scalar and tensor-tensor-scalar interactions in GMG, similar to those found in the scalar-tensor theories. The tensor-tensor-scalar interactions are important to estimate the amplitude of scalar perturbations induced by the tensor perturbation. The dominant contribution for the tensor-scalar-scalar interactions is the 4-derivative terms, which are enhanced by $\omega/m$. It is possible to fine-tune parameters, $\alpha_3 + \alpha_4=0$, to eliminate this contribution as shown in Eq.~(\ref{eq:coefcombined}). In this case, the next-to-leading order contribution comes from 3-derivative terms. These terms are not enhanced by $\omega/m$, but still give $\mathcal{O}(1)$ contributions to the gradient term and can give rise to the gradient instability. However, unlike in scalar-tensor theories, the coefficients of tensor-scalar-scalar interactions have no proportionality relation to those of tensor-tensor-scalar interactions in GMG. This implies that even after tuning model parameters to suppress tensor-scalar-scalar interactions, tensor-tensor-scalar interactions are not suppressed. In addition, in GMG, there are infinite higher order interactions of the form given by Eq.~(\ref{eq:higher}). Unless there are cancellations, these higher order terms are of the same order of magnitude as the cubic terms. This makes it difficult to predict the fate of the gradient instability inferred from the cubic interactions. These infinite series of interactions arise from the graviton potential which has a square-root form.

In GMG, we have a similar situation for self-interactions of scalar perturbations. Again due to  the square-root nature of the graviton potential, when we perturb the St\"uckelberg field, we have an infinite series of scalar interactions. For a spherically symmetric source, we can deal with the matrix square-root and it is possible to obtain the interactions in a closed form. It has been shown that we have an efficient Vainshtein mechanism from these self-interactions at least in a minimal model of GMG, and the scalar field perturbations are highly suppressed compared with the linear solution inside the Vainshtein radius \cite{Gumrukcuoglu:2021gua}. The situation is more complicated for tensor-scalar interactions. In this case, we cannot ignore the time dependence of perturbations and we cannot assume a simple symmetry to deal with the square-root of matrix.  It is conceivable that these infinite series of scalar self-interactions and scalar-tensor interactions suppress the growth of scalar perturbations due to the gradient instability inferred from the truncation of these interactions at cubic order. However, to confirm this expectation, we will need to find a way to deal with these infinite series of interactions for time dependent non-spherically symmetric perturbations. 

In dark energy models, there are three effects from tensor-scalar interactions: the quantum effects due to perturbative \cite{Creminelli:2018xsv} and non-perturbative \cite{Creminelli:2019nok} decay of the gravitational wave into scalars; and the classical instability of the scalar modes \cite{Creminelli:2019kjy}. The gravitational wave decays rely on the availability of a state with slower sound speed. In massive gravity models with self-acceleration, the scalar mode exclusively propagates at a speed larger than the speed of light, so the quantum decays are kinematically forbidden. On the other hand, the classical instability is a possibility, the potentially strong interactions prevent us to reach a definitive conclusion on this point. 

Since our benchmark $\Lambda_3$ theory, i.e. GMG, has also dynamical vector modes, it is a fair question to ask whether the vector sector can be destabilised in a similar way to the scalar sector. Our calculation of scalar-tensor interactions provides the clues for the vector modes. For instance, in the presence of gravitational waves, the vector modes can indeed be sourced by terms of the form $h_{ij} h_{jk} \partial_k \Pi_i$. However, to provide a high order contribution to the vector gradient that can cause an instability, an interaction term should have all of the following: (i) two spatial derivatives of the vector modes; (ii) one copy of the tensor mode; (iii) additional time or space derivative(s) such that it is enhanced by the gravitational wave frequency, thus compete with the canonical gradient term. For example, an interaction like $\dot{h}_{ij} \partial^i \Pi^k \,\partial^j\Pi_k$ would lead to a gradient instability. If such a term exists, it should also have a scalar analogue under the replacement $\Pi_i \to \partial_i\Pi$, which would correspond to a $5$-derivative scalar-tensor interaction at cubic order. However, examining \eqref{eq:Lag_tss} and \eqref{eq:cubicTSS}, we observe that the highest derivative degree of cubic scalar-tensor interactions is $4$, which would simply correspond to a mass term for the vector case. We therefore argue that the vector sector is not subject to an instability triggered by the tensor mode at cubic order.

An unsettling concern is the closeness of the LIGO energies to the strong coupling scale $\Lambda_3$ in dark energy models. It is therefore not clear whether the effects from scalar-tensor interactions can be employed to put bounds on the effective field theory parameters. However, an observation from LISA\footnote{https://lisa.nasa.gov/} can potentially evade this problem. LISA will be able to probe frequencies between 0.1 mHz and 1 Hz compared to LIGO's frequency of 10 Hz to 1000Hz, which is 4 orders of magnitude smaller than LIGO. Therefore we expect that LISA gravitational waves would bring more reliable constraints on dark energy models. Moreover, the high order interactions of the form $h (\partial\partial\Pi)^{n-1}$ can potentially be suppressed if the amplitude satisfies $h_0 < (m/\omega_{LISA})$.

\acknowledgments
For the purpose of open access, the authors have applied a Creative Commons Attribution (CC BY) licence to any Author Accepted Manuscript version arising from this work.  Supporting research data are available on reasonable request from the corresponding author. 

AEG is supported by a Dennis Sciama Fellowship at the University of Portsmouth. KK is supported by the UK Science and Technology Facilities Council (grant numbers ST/S000550/1 and ST/W001225/1).

\appendix

\section{Tensor-Tensor-Scalar interactions at cubic order}
\label{app:TTS}
In this Appendix, we present the expression for the interactions that involve two tensors and one scalar, before integrating out the non-dynamical degrees. We have
\begin{equation}
 \delta^{3}S \ni \frac{\Mpl^2}{2}\int a^3 dt\,d^3x\,\mathcal{L}^{tts}\,,
\end{equation}
where
\begin{align}
\mathcal{L}^{tts} =&\Phi\,\left[-\frac{1}{4}\,\dot{h}^{ij}\dot{h}_{ij}
-2\,H\,\dot{h}_{ij}h^{ij}
+\frac{1}{a^2}\,\partial^2h_{ij}\,h^{ij}
+\frac{3}{4\,a^2}\,\partial_ih_{jk}\,\partial^ih^{jk}
-\frac{1}{2\,a^2}\,\partial_ih_{jk}\,\partial^jh^{ik}
+\frac{m^2\xi\,(Q\,\xi-3\,J)}{4}\,h_{ij}h^{ij}
\right]
\nonumber\\
& +\frac{B}{a}\,\left[ -\partial^2\dot{h}_{ij}h^{ij}
+\partial_ih_{jk}\partial^k\dot{h}^{ij}
-\frac{1}{2}\,\partial^2h_{ij}\dot{h}^{ij}
-\frac{3}{2}\,\partial_ih_{jk}\partial^i\dot{h}^{jk}
\right]
\nonumber\\
&+\frac{m^2\xi}{4}\,\left[2\,J+Q\,\xi\,(2r-3)-G\,\xi^2(r-1)\right]h_{ij}h^{ij}\partial^2\Pi^L
-\frac{m^2\xi}{4}\,\left[3\,J+Q\,\xi\,(3r-5)-2\,G\,\xi^2(r-1)\right]h_{ij}h^{jk}\partial^i\partial_k\Pi^L
\nonumber\\
&
+\frac{m^2r\,\xi^2}{4}\,\left[H\,(6\,Q-G\,\xi)-3\,J_\chi+\xi\,Q_\chi\right]\,h_{ij}h^{ij}\,\Pi^0
+\frac{m^2r\,\xi}{2}\left(Q\,\xi-2\,J\right)\,\dot{h}_{ij}h^{ij}\,\Pi^0\,.
\label{eq:TTSterms-pre}
\end{align}
Substituting \eqref{eq:SHsolutions} into the above yields Eq.\eqref{eq:TTSterms}.

\section{Solving the constraints at next-to-leading order in gradient expansion}
\label{app:NTL-gradient}
In this Appendix, we relax the sub-horizon limit and allow next-to-leading order variations. These terms are normally suppressed, but in Sec.\ref{sec:TTS-interactions} we found that the leading order terms in the subhorizon approximation cancel. Therefore, we need to go to the next order. To this goal, we turn back to the equations of motion for the non-dynamical modes, given in Eqs.\eqref{eq:Bcons}-\eqref{eq:Pcons}.

We first solve Eq.\eqref{eq:Bcons} for $B$, obtaining:
\begin{equation}
B = -\frac{r^2}{a}\Pi^0+a\,\dot\Pi^L-\frac{2\,H\,(r+1)}{m^2a\,J\,\xi}\Phi\,.
\label{eq:Bsol-alt}
\end{equation}
We can also combine \eqref{eq:Fcons} with \eqref{eq:Pcons} to cancel the Laplacians of $\Phi$ (and accidentally, $B$) to obtain:
\begin{equation}
 \Phi = \xi\left(\frac{m^2J}{2\,H}+\frac{(2\,H\,Q-J_\chi)(r+1)}{J}\right)\Pi^0 - \frac{r}{3\,a^2\,H}\nabla^2\left[
\Pi^0 + \frac{a^2}{r}\,\dot\Pi^L - \frac{a^2\xi}{r}\left(\frac{m^2J\,r}{2\,H}+\frac{(2\,H\,Q-J_\chi)(r+1)}{J}\right)\Pi^L
\right]\,.
 \label{eq:Fsol-alt}
\end{equation}
We then use Eqs.\eqref{eq:Bsol-alt} and \eqref{eq:Fsol-alt} in the remaining equation to obtain
\begin{align}
&\left(\frac{m^2J}{2\,H}+\frac{2\,H\,Q-J_\chi}{J}\right)\left[-\frac{3\,m^2a^2J\,\xi}{2\,(r+1)}\,\Pi^0 +\nabla^2\left(\Pi^0 - \frac{m^2a^2J\,\xi}{2\,H\,(r+1)}\,\Pi^L\right)\right]\nonumber\\
&
\qquad\qquad\qquad-\frac{r}{3\,a^2H\,(r+1)\,\xi}\nabla^4\left[
\Pi^0 + \frac{a^2}{r}\,\dot\Pi^L - \frac{a^2\xi}{r}\left(\frac{m^2J\,r}{2\,H}+\frac{(2\,H\,Q-J_\chi)(r+1)}{J}\right)\Pi^L
\right]=0\,.
\end{align}
Assuming that $\Pi^L$ and $\dot{\Pi}^L$ do not get super-horizon corrections, we now look for solutions of the form 
\begin{equation}
\Pi^0 = \Pi^0_{(0)} + \nabla^{-2}\Pi^0_{(1)} + \mathcal{O}(\nabla^{-4})\,,  
\end{equation}
where the leading order term $\Pi^0_{(0)}$ is simply the sub-horizon limit solution presented in the third line of Eq.\eqref{eq:SHsolutions}. For $\Pi^0_{(1)}$, we find:
\begin{equation}
\Pi^0_{(1)} = \frac{3\,a^4H\,\xi}{r^2}\left(\frac{m^2J}{2\,H}+\frac{2\,H\,Q-J_\chi}{J}\right)\,\left[\left(\frac{m^2J\,r^2}{2\,H}+\frac{(2\,H\,Q-J_\chi)(1+r)^2}{J}\right)\xi\,\Pi^L-(r+1)\,\dot\Pi^L\right]\,.
\label{eq:P0-NTL}
\end{equation}
With this solution, we can combine \eqref{eq:Fcons} with \eqref{eq:Bcons}, then using the next-to-leading order term in Eq.\eqref{eq:P0-NTL}, we find the first order correction to 
\begin{equation}
\Phi=\Phi_{(0)}+\nabla^{-2}\Phi_{(1)}+\mathcal{O}(\nabla^{-4})\,,
\end{equation}
as
\footnote{Alternatively, one can instead use the gradient series of $\Pi^0$ in \eqref{eq:Fsol-alt} to solve for next-to leading order terms of $\Phi$. However, the $\mathcal{O}(\nabla^2)$ term in this equation vanishes at leading order, so one would need to calculate the second order $\Pi^0_{(2)}$ term to calculate the first order $\Phi_{(1)}$.}
\begin{equation}
\Phi_{(1)} = -\frac{3\,m^2J\,a^4\xi^2}{2}\,\left(\frac{m^2J}{2\,H} + \frac{2\,H\,Q-J_\chi}{J}\right)
\left[
\left(\frac{m^2J\,(r-1)}{2\,H} + \frac{(2\,H\,Q-J_\chi)(r+1)}{J}\right)\xi\,\Pi^L-\dot\Pi^L
\right]\,,
\end{equation}
with $\Phi_{(0)}$ given in the second line of Eq.\eqref{eq:SHsolutions}. 
Finally, we use the gradient expansions for $\Pi^0$ and $\Phi$ in Eq.\eqref{eq:Bsol-alt} to determine the first order correction to
\begin{equation}
 B = B_{(0)}+\nabla^{-2}B_{(1)}+\mathcal{O}(\nabla^{-4})\,,
\end{equation}
as
\begin{equation}
B_{(1)} = -\frac{3\,m^2a^3J\,\xi^2}{2}\,\left(\frac{m^2J}{2\,H}+\frac{2\,H\,Q-2\,J_\chi}{J}\right)\Pi^L\,,
\end{equation}
with $B_{(0)}$ given in the first equation of Eq.\eqref{eq:SHsolutions}. 

The result indicates that the $\mathcal{O}(\nabla^{-2})$ terms in $\Phi$ and $\Pi^0$ can decrease the order of derivatives by at least $1$, while from $B$ the decrease is always of $2$-derivative orders. For instance, inspecting the tensor-tensor-scalar interactions before replacing the non-dynamical modes, presented in Eq.\eqref{eq:TTSterms-pre}, we see that terms that include $B$ have already 3 derivatives while $\Phi$ and $\Pi^0$ appear in terms with at most 2 derivatives. We thus conclude that none of the next-to-leading order terms found in this Appendix contributes to the $2$-derivative terms in the final action Eq.\eqref{eq:TTSterms}.

\end{document}